\documentclass[aps,prd,twocolumn,showpacs,showkeys,amsmath,amssymb]{revtex4}
\usepackage{amsfonts}
\usepackage{amsmath}
\usepackage{graphicx}
\usepackage{subfigure}
\usepackage{dcolumn}
\usepackage{bm}
\usepackage{booktabs}
\usepackage[utf8]{inputenc}
\usepackage{autobreak}
\usepackage{float}

\usepackage{graphicx,graphics,dcolumn,booktabs,bm}
	\graphicspath{{figs/}}
\usepackage{float}
\usepackage{longtable,lscape}
\usepackage{txfonts}
\usepackage{overpic}
\usepackage{amssymb}
\usepackage{indentfirst}
\usepackage{epsfig}
\usepackage{feynmf}   
\usepackage{epstopdf}   
\usepackage{slashed}  
\usepackage{multirow}

\usepackage[colorlinks, citecolor=blue,anchorcolor=red,menucolor=red, linkcolor=red,filecolor=red,runcolor=red,urlcolor=blue,frenchlinks=red]{hyperref}

\usepackage{hhline}
\allowdisplaybreaks[3]
\usepackage{ulem}
\usepackage[usenames,dvipsnames]{color}

\makeatletter
\newcommand{\figcaption}{\def\@captype{figure}\caption}
\newcommand{\tabcaption}{\def\@captype{table}\caption}

\newcommand{\Rmnum}[1]{\expandafter\@slowromancap\romannumeral #1@}
\def\hlinewd#1{%
  \noalign{\ifnum0=`}\fi\hrule \@height #1 \futurelet
   \reserved@a\@xhline}
\makeatother

\newcommand\dqq{\left< \bar q q \right>}
\newcommand\duu{\left< \bar u u \right>}
\newcommand\ddd{\left< \bar d d \right>}
\newcommand\dss{\left< \bar s s \right>}
\newcommand\dqGq{\left< g_s \bar q \sigma G q \right>}
\newcommand\dsGs{\left< g_s \bar s \sigma G s \right>}
\newcommand\dGG{\left< g_s^2 GG \right>}
\newcommand\ms{m_s}
\newcommand\unit{\mathrm{GeV}}

\newcommand\Msq{M_B^2}
\newcommand\M{M_B}

\newcommand\mycite{\cite}

\begin{document}


\title{Mass spectra of $N\Omega$ dibaryons in the $^{3}S_1$ and $^{5}S_2$ channels}

\author{Xiao-Hui Chen$^1$}
\author{Qi-Nan Wang$^1$}
\author{Wei Chen$^1$}
\email{chenwei29@mail.sysu.edu.cn}
\author{Hua-Xing Chen$^2$}
\email{hxchen@seu.edu.cn}
\affiliation{$^1$School of Physics, Sun Yat-Sen University, Guangzhou 510275, China
\\
 $^2$School of Physics, Southeast University, Nanjing 210094, China }

\begin{abstract}
We study the mass spectra of the $N\Omega$ dibaryons in the $^{3}S_1$ and $^{5}S_2$ channels with $J^{P}=1^{+}$ and $2^{+}$ respectively, by using the method of QCD sum rules. We construct two dibaryon interpolating currents in the molecular picture and calculate their correlation functions and spectral densities up to dimension-16 condensates. Our results indicate that there may exist an $N\Omega$ dibaryon bound state in the $^{5}S_2$ channel with a binding energy of about $21\ \mathrm{MeV}$. The masses of the $^{3}S_1$ $N\Omega$ dibaryons with $J^{P}=1^{+}$ are predicted to be higher than the $N\Omega$ and $\Lambda\Xi$ thresholds, and thus can decay into these final states directly in S-wave. The $N\Omega (^{5}S_2)$ dibaryon bound state can decay into the octet-octet final states $\Lambda\Xi$ and $\Sigma\Xi$ in D-wave via the quark rearrangement mechanism. The existence of these $N\Omega$ dibaryons may be identified in the relativistic heavy-ion collision experiments in the future.
\end{abstract}

\keywords{QCD sum rules, molecular state, dibaryon} 
\pacs{12.39.Mk, 12.38.Lg, 14.40.Lb, 14.40.Nd}

\maketitle

\section{Introduction} 
Quantum chromodynamics (QCD) has been proven to be the fundamental theory of strong interaction between quarks and gluons, although it is still ambiguously at the low-energy region. The study of multiquarks could enrich our knowledge of hadron structures and hadron interactions, and provide more information about the nonperturbative behavior of QCD.
In the past decades, there have been impressive progress on the search for multiquark states, especially after the observations of hidden-charm XYZ and $P_c$ states~\cite{Liu:2019zoy,2016-Chen-p1-121,2016-Esposito-p1-97,2016-Richard-p1185-1212,2017-Ali-p123-198,2017-Lebed-p143-194,2018-Guo-p15004-15004,2018-Olsen-p15003-15003}.
	
Dibaryons are bound states of two color-singlet baryons with $B=2$.
The deuteron is the first dibaryon as a loosely bound state of a proton and a neutron observed in 1932~\cite{deuteron}. Until recent years, $d^*(2380)$ as another candidate was discovered in the two-pion production reaction $pn \to d\pi^0\pi^0$ and confirmed in $pn$ scattering by the COSY/CELSIUS and WASA-at-COSY Collaborations~\cite{Bashkanov:2008ih, 2011-Adlarson-p242302-242302,2013-Adlarson-p229-236,2014-Adlarson-p202301-202301,2015-Adlarson-p325-332}. The observation of such a resonance has inspired a lot of theoretical studies on its structure, such as the $\Delta\Delta$ bound state with the $I(J^P)=0(3^+)$ assignment proposed in Refs.~\cite{1999-Yuan-p45203-45203,2013-Gal-p172301-172301,2015-Chen-p25204-25204,2014-Huang-p34001-34001,2015-Huang-p71001-71001,dibaryon-sextet}. Besides, this structure was explained by a triangle singularity without invoking a dibaryon in Refs.~\cite{Molina:2021bwp,Ikeno:2021frl}. There the authors also explained why the peak is not observed in other reactions where it has been searched for. Moreover, the famous H state was first predicted by Jaffe as a genuine bound state with strangeness $S=-2$ well below the $\Lambda\Lambda$ threshold~\cite{H-dibaryon}. Although numerous efforts have been done in the past several decades, there has been no any evidence for the existence of such an H dibaryon to date~\cite{dibaryon-review}.

Besides the above octet-octet systems, the $N\Omega$ dibaryon in the octet-decuplet representation is also expected to have a bound state~\cite{Goldman:1987ma,Li:1999bc,Pang:2003ty,Chen:2011zzb,Sekihara:2018tsb,Xiao:2020alj,Etminan:2014tya,Iritani:2018sra}, since the Pauli exclusion principle does not operate and the color-magnetic interaction is attractive in this system. For the S-wave $N\Omega$ dibaryons, they can be in $^{3}S_1$ and $^{5}S_2$ (symbol $^{2S+1}L_{J}$, where $S, L, J$ are the total spin, relative orbit angular momentum and total angular momentum, respectively) channels with $I(J^P)=\frac{1}{2}(1^+)$ and $\frac{1}{2}(2^+)$, respectively. The couplings of the $^{5}S_2$ $N\Omega$ dibaryon to the S-wave octet-octet channels, e.g., $\Lambda\Xi$, $\Sigma\Xi$, are strongly suppressed in D-wave. However, the $^{3}S_1$ $N\Omega$ dibaryon can strongly couple to the S-wave octet-octet channels in S-wave. 
In Ref.~\cite{Goldman:1987ma}, an $N\Omega$ state with $S=-3,\, I=1/2,\, J=2$ was first proposed as a deeply bound dibaryon candidate with binding energy $E_{N\Omega}=140$ MeV and 250 MeV in two different quark models.
Under the chiral SU(3) constituent quark model, the same $N\Omega$ bound state was suggested with much smaller binding energy ranging from 3.5 MeV to 12.7 MeV~\cite{Li:1999bc}. This dibaryon was also studied in the extended quark delocalization and color screening model, in which the binding energy was obtained around 62 MeV~\cite{Pang:2003ty}. In Ref.~\cite{Chen:2011zzb}, a chiral quark model and a quark delocalization color screening model were employed to calculate the baryon-baryon scattering phase shifts to look for dibaryon resonances. For the $N\Omega$ system, they found a bound state about 53 MeV lower than the $N\Omega$ threshold in the quark delocalization color screening model, while no bound state exist in the chiral quark model. 
Based on a baryon-baryon interaction model with meson exchanges, a quasibound state of $N\Omega(^5S_2)$ 
was obtained with the binding energy 0.1 MeV and the decay width 1.5 MeV mainly into $\Lambda\Xi$ mode~\cite{Sekihara:2018tsb}. Such D-wave decay properties of $d_{N\Omega}\to\Lambda\Xi$ and $d_{N\Omega}\to\Sigma\Xi$ were recently studied in a phenomenological Lagrangian approach and their result showed that the total decay width of the $N\Omega(^5S_2)$ dibaryon bound state was in the range of a few hundred 
keV~\cite{Xiao:2020alj}.

The HAL QCD Collaboration also studied the $N\Omega$ interaction in the $^{5}S_2$ channel in (2+1)-flavor lattice QCD simulations~\cite{Etminan:2014tya}. By employing the quark masses corresponding to $m_{\pi}=875$ MeV and $m_K=916$ MeV, they found that the interaction between $N$ and $\Omega$ was attractive in all distances, and the binding energy is $B_{N\Omega}=18.9$ MeV. Recently, they improved their LQCD simulations near the physical points ($m_{\pi} \sim 146\ \mathrm{MeV}$, $m_K\sim525\ \mathrm{MeV}$)~\cite{Iritani:2018sra} and found that the interaction between the two baryons is still attractive in all distances. A shallow quasi-bound state below the $N\Omega$ threshold was found having much smaller binding energy $B_{N\Omega}=1.54$ MeV (2.46 MeV) without (with) the Coulomb attraction. 

In Ref.~\cite{Morita:2016auo}, the proton-omega two-particle momentum correlation function in relativistic heavy ion collisions was 
studied. They extracted the strong $p\Omega$ attractive interaction for the spin-2 channel by measuring the ratio of correlation functions between small and large collision systems. Experimentaly, such $p\Omega$ correlation function in relativistic heavy ion collisions was also measured by the STAR experiment at RHIC~\cite{STAR:2018uho}. The measured ratio of the correlation function slightly favored a $p\Omega$ bound state with a binding energy of $\sim 27$ MeV. 

In this work, we shall study the existence of the $N\Omega$ dibaryons in both $^{3}S_1$ and $^{5}S_2$ channels by using the method of QCD sum rules, following the previous investigation of the $\Omega\Omega$ system~\cite{Chen:2019vdh}. This paper is organized as follows. In Sec. II, we introduce the formalism of QCD sum rules and construct the dibaryon interpolating currents with $J^P=1^+, 2^+$. In Sec. III, we perform the numerical analyses for all channels and predicted their masses and coupling constants. The last section is a short summary. 

%
%
%

\section{QCD Sum Rules for Dibaryon Systems}

QCD sum rule is a powerful non-perturbative approach to investigate the hadron properties, such as the hadron masses, decay widths, magnetic moments and so on~\cite{Shifman:1978bx, QCD-sum-rule-Reinders}. To establish the dibaryon QCD sum rules, we first need to construct the interpolating currents for the $N\Omega$ system. We shall use the two local Ioffe currents to represent the nucleon~\cite{1981-Ioffe-p317-341,1983-Ioffe-p67-67}
\begin{align}
	J^N_1(x) &= \epsilon^{abc} \left[ u^T_a(x) C\gamma_\mu u_b(x) \right] \gamma^\mu \gamma^5 d_c(x), 
	\\
	J^N_2(x) &= \epsilon^{abc} \left[ u^T_a(x) C\gamma_5 d_b(x) \right] u_c(x),
\end{align}
and the current for $\Omega$ baryon is written as
\begin{equation}
J_\mu^\Omega(x)=\epsilon^{abc}\left[s^T_a (x) C\gamma_\mu s_b (x) \right]s_c (x),
\end{equation}
in which $u,d,s$ represent the up, down and strange quark field respectively, $a, b, c$ are the color indices, $C$ is the charge conjugation matrix and $T$ the transpose operation. Note that the currents $J^N_1(x)$ and $J^N_2(x)$ contain the quark configuration of proton, and the current of neutron can be obtained by the replacement $u \leftrightarrow d$. However, such difference does not affect the sum rule analyses since we shall work in the SU(2) isospin symmetry without considering the effect of isospin breaking in the following calculation.

In the molecular picture, we can construct two $N\Omega$ dibaryon interpolating currents as
\begin{align}
	\begin{split}
	J^{N\Omega}_{\mu\nu,\,1}(x) &= \epsilon^{abc}\epsilon^{def} \left[ u^T_a(x) C\gamma_\lambda u_b(x) \right] \left( \gamma^\lambda \gamma^5 d_c(x) \right)^T \cdot 
	\\ &\quad
	 C\gamma_\mu \cdot s_f(x) \left[ s^T_d(x) C\gamma_\nu s_e(x) \right],
	 \label{J1}
	\end{split}
	\\
	\begin{split}
	J^{N\Omega}_{\mu\nu,\,2}(x) &= \epsilon^{abc}\epsilon^{def} \left[ u^T_a(x) C\gamma_5 d_b(x) \right] u^T_c(x) \cdot 
	\\ &\quad
	C\gamma_\mu \cdot s_f(x) \left[ s^T_d(x) C\gamma_\nu s_e(x) \right].
	\label{J2}
	\end{split}
\end{align}
Both of these two interpolating currents can couple to the $^{3}S_1$ and $^{5}S_2$ channels with $I(J^P)=\frac{1}{2}(1^+)$ and $\frac{1}{2}(2^+)$ respectively. The two-point correlation functions induced by the above currents are
\begin{eqnarray} 
\Pi_{\mu\nu,\,\rho\sigma}(q^2) &=& i \int\mathrm{d^4} x \  \mathrm{e}^{iq\cdot x} \left< 0 \left| \mathrm{T} \left\{ J^{N\Omega}_{\mu\nu}(x)  J^{N\Omega\dag}_{\rho\sigma}(0) \right\} \right| 0 \right>.
\label{tensorCF}
\end{eqnarray}

In general, the correlation function $\Pi_{\mu\nu,\,\rho\sigma}(q^2)$ contains several different invariant functions $\Pi(q^2)$ referring to different, spin-1 and spin-2 hadron states. One can pick out these invariant functions by using the projectors introduced in Ref.~\cite{2014-Chen-p201-215,2017-Chen-p114005-114005,2017-Chen-p114017-114017}.  For the $J^P=1^+$ and $2^+$ pieces, the corresponding invariant functions $\Pi_{1}(q^2)$ and $\Pi_{2}(q^2)$ can be extracted via the following projectors 
\begin{align}
\notag
P_{1A}&=\left[\eta_{\mu\rho}\eta_{\nu\sigma}-(\rho \leftrightarrow \sigma) \right]\, , &~\mbox{for}\, J^P=1^+,\\
P_{2S}&=\eta_{\mu\rho}\eta_{\nu\sigma}+\eta_{\mu\sigma}\eta_{\nu\rho}-\frac{2}{3}\eta_{\mu\nu}\eta_{\rho\sigma}\, , &~\mbox{for}\, J^P=2^+, \label{projectors}
\end{align}
where the $\eta_{\mu\nu}$ is defined as 
\begin{equation}
\begin{split}
\eta_{\mu\nu}&=\frac{q_\mu q_\nu}{q^2}-g_{\mu\nu}.
\end{split}
\end{equation}
	
At the QCD side, the correlation functions can be calculated via the operator product expansion (OPE) method as a series of various QCD condensates, such as the quark condensate, gluon condensate, quark-gluon mixed condensate and some other higher dimension condensates, parameterize the QCD non-perturbative effect. In this work, the correlation functions shall be calculated up to dimension-16 condensates at the leading order of $\alpha_s$.

At the hadronic level, the invariant function $\Pi(q^2)$ can also be  expressed as a dispersion relation 
\begin{equation}
\Pi\left(q^{2}\right)=\left(q^{2}\right)^{N}\int_{0}^{\infty}\mathrm{d}s\frac{\rho_{phen}\left(s\right)}{s^{N}\left(s-q^{2}-\mathrm{i}\epsilon\right)}+\sum_{k=0}^{N-1}b_{n}\left(q^{2}\right)^{k}\,
, \label{eq:dispersion relation}
\end{equation}
where $b_n$ is an unknown subtraction constant. The spectral density $\rho_{phen}(s)$ can be usually obtained by inserting intermediate states $|X\rangle$ of which the quantum numbers are the same as the interpolating current. We adopt the duality ansatz ``one narrow resonance + continuum'' 
as the spectral function parametrization
\begin{align}
	\notag
	\rho_{phen}(s) &\equiv \frac{ \mathrm{Im}\Pi(s) }{\pi} =
	\sum_X \delta (s-m_{X}^{2}) \langle 0|J|X \rangle \langle X|J^{\dagger}|0 \rangle
	\\
	&=f_L^2\delta(s-m_L^2)+\theta(s-s_0) \rho_H(s),
	\label{eq:spectral density}
\end{align}
where the $\delta(s-m_L^2)$ function describes the lowest-lying hadron state and the second part contains the contributions from excited states and continuum. A continuum threshold $s_0$ is introduced here to describe the cutoff between the lowest-lying resonance and continuum. The coupling constants for axial-vector and tensor states are defined as
\begin{align}
\langle0|J^{N\Omega}_{\mu\nu}|X_A\rangle&=f_{A}\varepsilon_{\mu\nu\alpha\beta}\epsilon^{\alpha}q^{\beta}, 
\label{vectorcoupling}
\\
\langle0|J^{N\Omega}_{\mu\nu}|X_T\rangle&=f_{T} \epsilon_{\mu\nu},
\label{tensorcoupling}
\end{align}
in which $\epsilon^{\alpha}$ and $\epsilon_{\mu\nu}$ are the corresponding polarization vector and tensor, and $\varepsilon_{\mu\nu\alpha\beta}$ is the fully symmetric tensor.

Due to the quark-hadron duality, the correlation functions obtained at the QCD side and phenomenological side can be equal to each other. To suppress the contributions from excited states and continuum, and also to remove the unknown subtraction constants $b_n$ in Eq.~\eqref{eq:dispersion relation}, one usually performs the Borel transform to the correlation functions at both sides
\begin{equation}
\Pi(s_0,\, M_B^2)=\int_{t_c}^{s_0}  \rho(s) e^{-s/M_B^2} \,\mathrm{d}s= f_L^2 e^{ -m_L^2/M_B^2 },
\label{sumrules}
\end{equation}
in which the spectral function in Eq.~\eqref{eq:spectral density} is considered, and $t_c=(3m_s)^2$ is the physical threshold. The parameter $M_{B}$ is the Borel mass introduced by Borel transform, which is an unphysical parameter and should be irrelevant to the hadron mass. According to Eq.~\eqref{sumrules}, the lowest-lying hadron mass can be extracted as 
\begin{equation}
	m_L^2\left(s_0,\, M_B^2\right)=\frac{ \int_{t_c}^{s_0}  s\cdot\rho(s) e^{-s/M_B^2}\,\mathrm{d}s }
	{ \int_{t_c}^{s_0}  \rho(s) e^{-s/M_B^2}\,\mathrm{d}s }.
	\label{hadronmass}
\end{equation}
The spectral density $\rho(s)$ for the interpolating currents $J^{N\Omega}_{\mu\nu,\,1}(x)$ and $J^{N\Omega}_{\mu\nu,\,2}(x)$ are calculated and listed in the Appendix~\ref{spectraldensities} since these expressions are complicate and lengthy to show here.

\section{Analyses for dibaryon systems}
We shall apply the following values for the various QCD condensates and quark masses in the following numerical analyses~\mycite{PDG, condensates1, condensates2, condensates3, condensates4, condensates5, condensates-chenhuaxing,QCD-sum-rule-chenhuaxing}
\begin{gather}
\nonumber
\duu=\ddd=\dqq= -(0.24\pm0.03)^3 \,\unit^3,  \\ 
\nonumber
\dss=-(0.8\pm0.1) \times (0.24\pm0.03)^3 \,\unit^3,  \\ 
\nonumber
\dGG=(0.48 \pm 0.14) \, \unit^4, \\
\dqGq=-M_0^2 \dqq,   \label{QCDparameters} \\ 
\nonumber
\dsGs= -M_0^2 \dss,  \\ 
\nonumber
M_0^2= (0.8 \pm 0.2) \, \unit^2,  \\ 
\nonumber
m_s= 95 ^{+9} _{-3} \, \mathrm{MeV}, 
\end{gather}
in which we don't distinguish the up and down quarks in the SU(2) chiral symmetry and $m_u=m_d=m_q=0$.

As mentioned above, the tensor correlation function $\Pi_{\mu\nu,\,\rho\sigma}(q^2)$ can couple to both $J^P=1^+$ and $2^+$ diabryon states. We shall study the $N\Omega$ dibaryon with $J^P=2^+$ at first,  by using the interpolating current $J^{N\Omega}_{\mu\nu,\,1}(x)$ in Eq.~\eqref{J1} as an example. 

\begin{figure}[ht]
\centering
\includegraphics[width=0.48\textwidth]{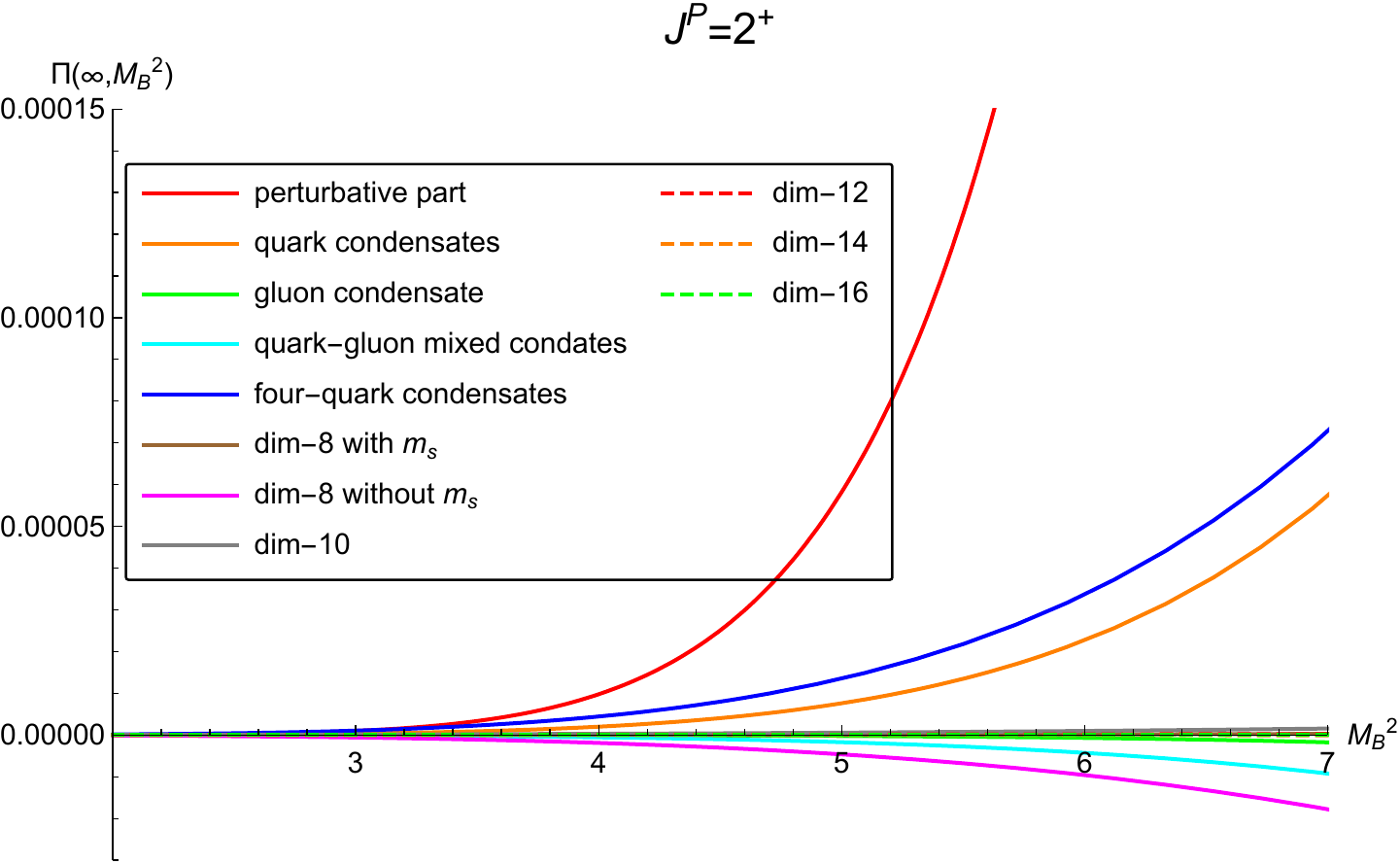}
\caption{OPE convergence in the $J^P=2^+$ tensor channel for the interpolating current $J^{N\Omega}_{\mu\nu,\,1}(x)$.}
\label{fig:J1&2+@ope}
\end{figure}

It is shown that the correlation function in Eq.~\eqref{sumrules} is the function of the continuum threshold $s_0$ and the Borel mass $M_B$.
To perform the QCD sum rule analysis, one should determine the suitable working regions for these two parameters. The lower bound on the Borel mass $M_B^{2}$ can be obtained by ensuring the OPE convergence of the correlation function. For this purpose, we investigate the tensor correlation function for the current $J^{N\Omega}_{\mu\nu,\,1}(x)$
\begin{align}
\begin{autobreak}
\Pi(\infty,\,\Msq)=
1.49 \times 10^{-10} \M^{16}
+4.72 \times 10^{-10} \M^{12}
+3.78 \times 10^{-9} \M^{10}
-7.30 \times 10^{-9} \M^8
+4.51 \times 10^{-9} \M^6
+4.55 \times 10^{-9} \M^4
-8.06 \times 10^{-9} \M^2
+4.95 \times 10^{-10}\, ,
\end{autobreak}
\end{align}
where the continuum threshold $s_0$ tends to infinity and the parameter values in Eq.~\eqref{QCDparameters} are adopted. 

\begin{figure}[ht]
\centering
\includegraphics[width=0.48\textwidth]{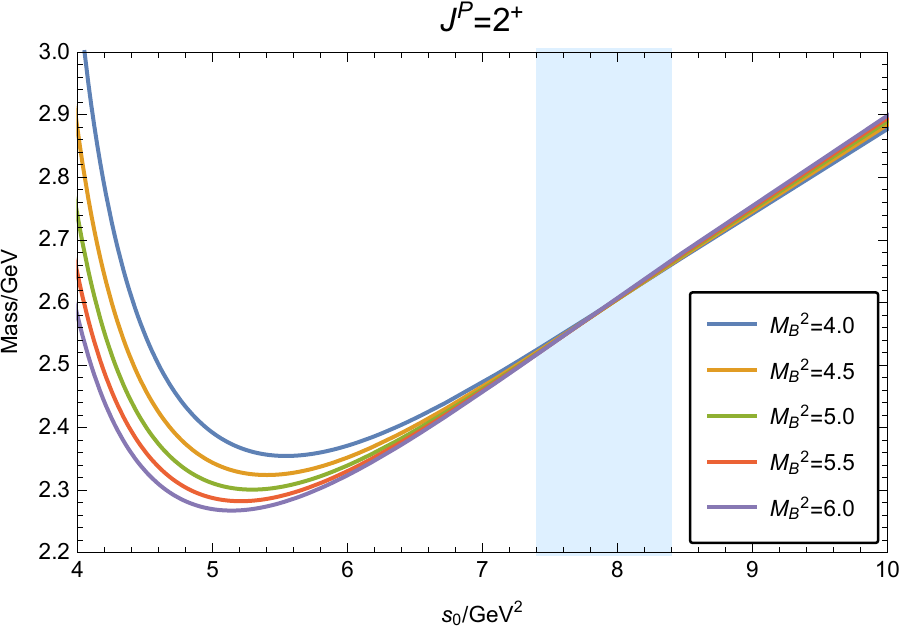}
\caption{Extracted hadron mass with respect to $s_0$ in the $J^P=2^+$ tensor channel for the interpolating current $J^{N\Omega}_{\mu\nu,\,1}(x)$.}
\label{fig:J1&2+_s0}
\end{figure}

It is shown that the correlation function $\Pi(\infty,\,\Msq)$ is the polynomial of the Borel mass $M_B$ up to dimension sixteen. We plot the $\Pi (\infty,\, \Msq)$ term by term from the perturbative contribution to various nonperturbative condensate contribution in Fig.~\ref{fig:J1&2+@ope}. We can find that the contributions from quark condensates and four-quark condensates are much larger than those from any other condensates while $M_B^2$ is large enough, indicating they are the dominant nonperturbative contributions. To ensure the convergence of the OPE series, we require that the quark condensates and the four-quark condensates be less than one half of the perturbative term. The lower bound on the Borel mass is thus obtained as $M_{Bmin}^{2}= 4.0$ GeV$^2$.

\begin{figure}[ht]
\centering
\includegraphics[width=0.48\textwidth]{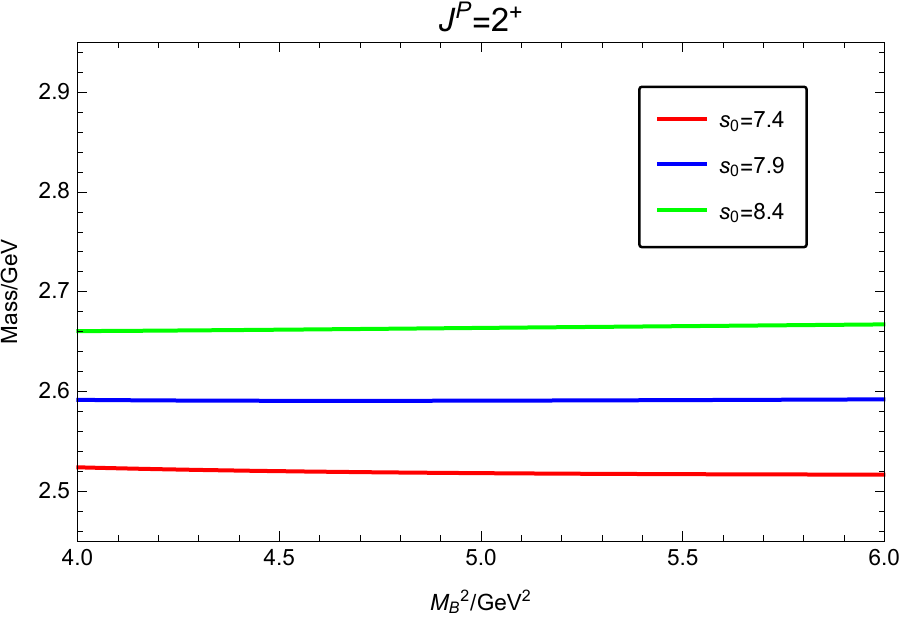}
\caption{Extracted hadron mass with respect to $M_B^2$ in the $J^P=2^+$ tensor channel for the interpolating current $J^{N\Omega}_{\mu\nu,\,1}(x)$.}
\label{fig:J1&2+_MB}
\end{figure}

We show the variation of the extracted hadron mass with respect to the continuum threshold $s_0$ with Borel mass $4.0 \,\text{GeV}^{2}<M_{B}^{2}< 6.0 \, \text{GeV}^{2}$ in Fig.~\ref{fig:J1&2+_s0}. In principle, the extracted hadron mass should not dependent to the unphysical parameter $M_B^2$. To eliminate such a dependence, we choose the working region of the continuum threshold as 7.4 GeV$^2 $$\leq s_0\leq 8.4$ GeV$^2$ in Fig.~\ref{fig:J1&2+_s0}.

In Fig.~\ref{fig:J1&2+_MB}, we give the Borel curves of the extracted hadron mass with respect to $M_B^2$, which reveals very good stability in the above parameter regions. The mass of the $N\Omega$ dibaryon with $J^P=2^+$ is then obtained as
\begin{align}
m_{N\Omega,\, 2^+}=(2.59\pm 0.17)\, \mbox{GeV}\, , \label{massTtensor}
\end{align}
and the corresponding coupling constant is
\begin{align}
f_{N\Omega,\, 2^+}=\left( 6.68\pm0.46 \right) \times 10^{-4}\, \mbox{GeV}^8\, .
\end{align}
where the errors come from the uncertainties of the Borel mass $M_B$, the threshold $s_0$ and the QCD parameters in Eq.~\eqref{QCDparameters}. The central value of the mass prediction is about 21 MeV below the $N{\Omega}$ two-baryon threshold~\cite{PDG}, which may suggest the existence of a loosely bound molecular state of the tensor $N\Omega$ dibaryon with $J^P=2^+$. The prediction of the binding energy is roughly agreement with 
the result of recent HAL QCD calculation near the physical points~\cite{Iritani:2018sra}.

\begin{figure}[H]
\centering
\includegraphics[width=0.48\textwidth]{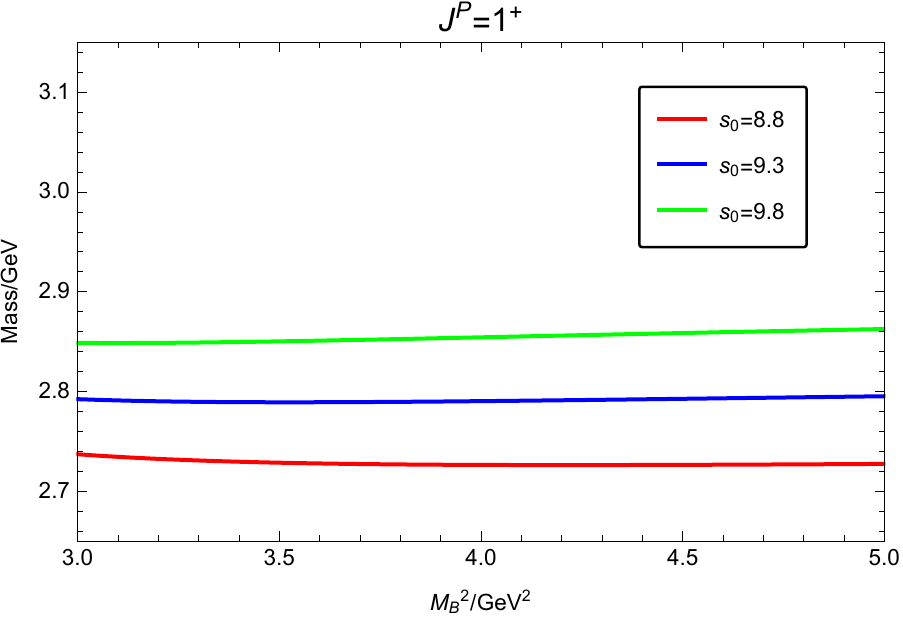}
\caption{Extracted hadron mass with respect to $M_B^2$ in the $J^P=1^+$ channel for the interpolating current $J^{N\Omega}_{\mu\nu,\,1}(x)$.}
\label{fig:J1&1+_MB}
\end{figure}

For the $J^P=1^+$ channel, we perform similar sum rule analyses and obtained the parameter working regions as 8.8 GeV$^2$ $\leq s_0\leq 9.8$ GeV$^2$, and 3.0 GeV$^2$ $\leq M_B^2 \leq 5.0$ GeV$^2$. Within these parameter regions, the mass curves are plotted in Fig.~\ref{fig:J1&1+_MB} 
with reasonable Borel stabilities. Finally, the hadron mass and the coupling constant for the $N\Omega$ dibaryon with $J^P=1^+$ can be extracted as  
\begin{align}
m_{N\Omega,\, 1^+}&=(2.79\pm 0.13)\, \mbox{GeV}\, ,\\
f_{N\Omega,\, 1^+}&=\left( 5.19\pm 1.20 \right) \times 10^{-4}\, \mbox{GeV}^8\, .
\end{align}
The obtained mass is much higher than the mass of tensor state in Eq.~\eqref{massTtensor}. This mass is above the $N\Omega$ and $\Lambda\Xi$ two-baryon thresholds, which means that the $N\Omega$ system can not form a bound state in the $^3S_1$ channel. 

\begin{figure}[ht]
\centering
\includegraphics[width=0.47\textwidth]{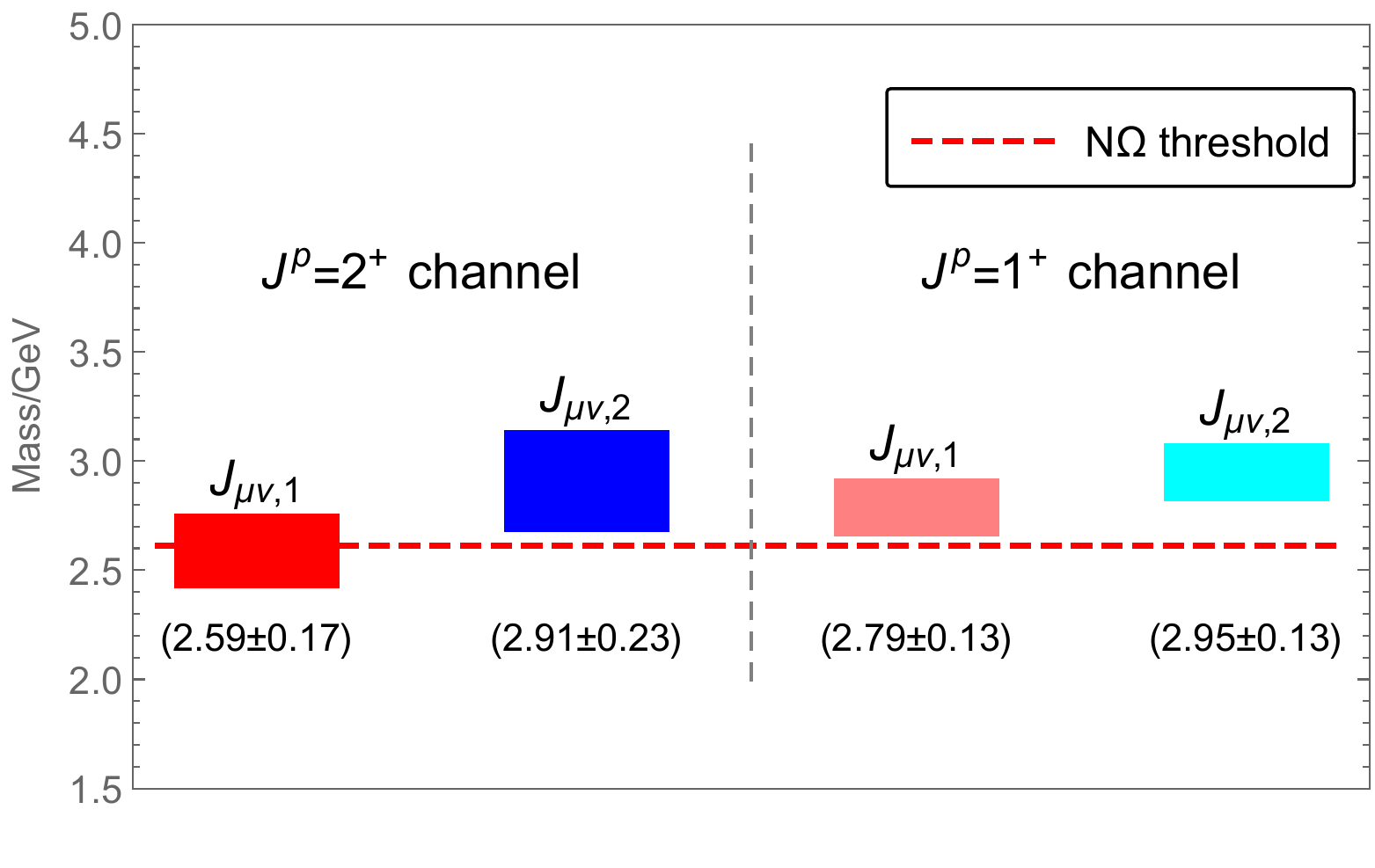}
\caption{Mass predictions for the $N\Omega$ states with $J^{P}=1^{+}$ and $2^{+}$, comparing with the $N\Omega$ two-baryon threshold.}
\label{fig:spectra}
\end{figure}

We also study the $N\Omega$ states by using the interpolating current $J^{N\Omega}_{\mu\nu,\,2}(x)$ in Eq.~\eqref{J2}, with the same procedure as the above analyses. The extracted hadron masses for the $N\Omega$ states with $J^{P}=1^{+}$ and $2^{+}$ are around 2.9 GeV, which is much higher than those obtained from the $J^{N\Omega}_{\mu\nu,\,1}(x)$. 
We collect the numerical results for the both the $J^{N\Omega}_{\mu\nu,\,1}(x)$ and $J^{N\Omega}_{\mu\nu,\,2}(x)$ in Table~\ref{NR} and compare with the $N\Omega$ threshold in Fig.~\ref{fig:spectra} for convenience. It shows that only one $N\Omega$ dibaryon with $J^P=2^+$ lies below the two-baryon threshold, and thus form a bound state.

\begin{table*}[ht]
\caption{Numerical results for the $J^P=1^+$ and $J^P=2^+$ channels in both $J^{N\Omega}_{\mu\nu,\,1}(x)$ and $J^{N\Omega}_{\mu\nu,\,2}(x)$.}
\renewcommand\arraystretch{1.7} 
\setlength{\tabcolsep}{1.em}{ 
\begin{tabular}{ccccccc}
\hline \hline
current & $J^P$	& $\text{mass}/\mbox{GeV}$ & $\text{coupling}/10^{-4}\,\mbox{GeV}^8$ & $s_0/\mbox{GeV}^2$ & $M_B^2/\mbox{GeV}^2$
\\ \hline
\multirow{2}*{ $J^{N\Omega}_{\mu\nu,1}$ } & $2^+$ & $2.59\pm0.17$ & $6.68\pm0.46$ & $[7.4,8.4]$ & $[4.0,6.0]$
\\ \cline{2-6}
& $1^+$ & $2.79\pm0.13$ & $5.19\pm1.20$ & $[8.8,9.8]$ & $[3.0,5.0]$
\\ \hline
\multirow{2}*{ $J^{N\Omega}_{\mu\nu,2}$ } & $2^+$ & $2.91\pm0.23$ & $5.58\pm1.16$ & $[9.2,10.2]$ & $[5.0,7.0]$
\\ \cline{2-6}
& $1^+$ & $2.95\pm0.13$ & $2.84\pm0.17$ & $[9.5,10.5]$ & $[4.0.6.0]$
\\ \hline \hline
\label{NR}
\end{tabular}
}
\end{table*}

\section{Summary}
In this work, we investigate the $N\Omega$ dibaryon systems with spin-parity $J^P=1^+, 2^+$, isospin $I=1/2$ and strangeness $S=-3$ in the framework of QCD sum rules. After constructing two tensor $N\Omega$ dibaryon interpolating currents in the molecular picture, we evaluate the correlation functions and the spectral densities up to dimension-16 nonperturbative condensates at the leading order of $\alpha_s$ via the operator product expansion method. 

We establish Laplace sum rules for the interpolating currents $J^{N\Omega}_{\mu\nu,\,1}(x)$ and $J^{N\Omega}_{\mu\nu,\,2}(x)$ in suitable 
working regions of the continuum threshold $s_0$ and Borel mass $M_B^2$. The hadron masses are then extracted reliably for both of them and four $N\Omega$ dibaryons are predicted. Among these four predicted states, the dibaryon with $J^P=2^+$ from $J^{N\Omega}_{\mu\nu,\,1}(x)$ lies about 
21 MeV below the $N\Omega$ threshold, implying the possible existence of a bound state. Within errors, this binding energy is roughly agreement with the prediction by recent LQCD calculation (2.46 MeV) near the physical points~\cite{Iritani:2018sra}, and the measurement of the $p\Omega$ correlation function by STAR Collaboration ($\sim 27$ MeV)~\cite{STAR:2018uho}. 

The other three dibaryon states predicted in Table~\ref{NR} are all above the $N\Omega$ and $\Lambda\Xi$ thresholds. All of them can directly decay into the $N\Omega$ final states in S-wave. For the two $N\Omega$ dibaryons in $^3S_1$ channel with $J^P=1^+$, they can also decay into the $\Lambda\Xi$ and $\Sigma\Xi$ final states with larger phase spaces in S-wave via the quark rearrangement mechanism. 
Although in D-wave, the $\Lambda\Xi$ channel was considered as the dominant decay mode for the $N\Omega$ dibaryons with $J^P=2^+$~\cite{Sekihara:2018tsb,Xiao:2020alj}. The $N\Omega  (^{5}S_2)\to\Lambda+\Xi+\pi$ process may be also an interesting decay channel for the weakly bound state of $N\Omega  (^{5}S_2)$ dibaryon, while the two-body $\Lambda\Xi, \Sigma\Xi$ decays are strongly suppressed in D-wave. These decay processes may be identified in the heavy-ion collision experiments like STAR, RHIC and J-PARC, where plenty of hyperons will be produced~\cite{Cho:2010db,Cho:2011ew}. Moreover, the measurement of the energy and width of the lowest $p\Omega^-$ atomic state in these experiments could be rather instructive, as it gives access to the $p\Omega^-$ scattering length, which would indicate that the hadronic $p\Omega^-$ system is almost bound or weakly bound~\cite{STAR:2018uho}. Recently, a new source of baryons and antibaryons has been proposed at a tau-charm factory like BESIII, or a super tau-charm factory like STCF or SCTF, where the numerous $J/\psi$ and $\psi^\prime$ event data will be able to supply copious hyperons such as $\Lambda, \Xi, \Omega$ and so on~\cite{Yuan:2021yks}.

\section*{ACKNOWLEDGMENTS}

This work is supported in part by National Key R$\&$D Program of China under Contracts No. 2020YFA0406400, the National Natural Science Foundation of China under Grants No. 11722540 and No. 12075019, and the Fundamental Research Funds for the Central Universities.


\begin{thebibliography}{47}
\expandafter\ifx\csname natexlab\endcsname\relax\def\natexlab#1{#1}\fi
\expandafter\ifx\csname bibnamefont\endcsname\relax
  \def\bibnamefont#1{#1}\fi
\expandafter\ifx\csname bibfnamefont\endcsname\relax
  \def\bibfnamefont#1{#1}\fi
\expandafter\ifx\csname citenamefont\endcsname\relax
  \def\citenamefont#1{#1}\fi
\expandafter\ifx\csname url\endcsname\relax
  \def\url#1{\texttt{#1}}\fi
\expandafter\ifx\csname urlprefix\endcsname\relax\def\urlprefix{URL }\fi
\providecommand{\bibinfo}[2]{#2}
\providecommand{\eprint}[2][]{\url{#2}}


%
  
%

\bibitem[{\citenamefont{Liu et~al.}(2019)\citenamefont{Liu, Chen, Chen, Liu,
  and Zhu}}]{Liu:2019zoy}
\bibinfo{author}{\bibfnamefont{Y.-R.} \bibnamefont{Liu}},
  \bibinfo{author}{\bibfnamefont{H.-X.} \bibnamefont{Chen}},
  \bibinfo{author}{\bibfnamefont{W.}~\bibnamefont{Chen}},
  \bibinfo{author}{\bibfnamefont{X.}~\bibnamefont{Liu}}, \bibnamefont{and}
  \bibinfo{author}{\bibfnamefont{S.-L.} \bibnamefont{Zhu}},
  \bibinfo{journal}{Prog. Part. Nucl. Phys.} \textbf{\bibinfo{volume}{107}},
  \bibinfo{pages}{237} (\bibinfo{year}{2019}), \eprint{arXiv:1903.11976}.

\bibitem[{\citenamefont{Chen et~al.}(2016)\citenamefont{Chen, Chen, Liu, and
  Zhu}}]{2016-Chen-p1-121}
\bibinfo{author}{\bibfnamefont{H.-X.} \bibnamefont{Chen}},
  \bibinfo{author}{\bibfnamefont{W.}~\bibnamefont{Chen}},
  \bibinfo{author}{\bibfnamefont{X.}~\bibnamefont{Liu}}, \bibnamefont{and}
  \bibinfo{author}{\bibfnamefont{S.-L.} \bibnamefont{Zhu}},
  \bibinfo{journal}{Phys. Rept.} \textbf{\bibinfo{volume}{639}},
  \bibinfo{pages}{1} (\bibinfo{year}{2016}), \eprint{arXiv:1601.02092}.

\bibitem[{\citenamefont{Esposito et~al.}(2016)\citenamefont{Esposito, Pilloni,
  and Polosa}}]{2016-Esposito-p1-97}
\bibinfo{author}{\bibfnamefont{A.}~\bibnamefont{Esposito}},
  \bibinfo{author}{\bibfnamefont{A.}~\bibnamefont{Pilloni}}, \bibnamefont{and}
  \bibinfo{author}{\bibfnamefont{A.~D.} \bibnamefont{Polosa}},
  \bibinfo{journal}{Phys. Rept.} \textbf{\bibinfo{volume}{668}},
  \bibinfo{pages}{1} (\bibinfo{year}{2016}).

\bibitem[{\citenamefont{Richard}(2016)}]{2016-Richard-p1185-1212}
\bibinfo{author}{\bibfnamefont{J.-M.} \bibnamefont{Richard}},
  \bibinfo{journal}{Few Body Syst.} \textbf{\bibinfo{volume}{57}},
  \bibinfo{pages}{1185} (\bibinfo{year}{2016}).

\bibitem[{\citenamefont{Ali et~al.}(2017)\citenamefont{Ali, Lange, and
  Stone}}]{2017-Ali-p123-198}
\bibinfo{author}{\bibfnamefont{A.}~\bibnamefont{Ali}},
  \bibinfo{author}{\bibfnamefont{J.~S.} \bibnamefont{Lange}}, \bibnamefont{and}
  \bibinfo{author}{\bibfnamefont{S.}~\bibnamefont{Stone}},
  \bibinfo{journal}{Prog. Part. Nucl. Phys.} \textbf{\bibinfo{volume}{97}},
  \bibinfo{pages}{123} (\bibinfo{year}{2017}).

\bibitem[{\citenamefont{Lebed et~al.}(2017)\citenamefont{Lebed, Mitchell, and
  Swanson}}]{2017-Lebed-p143-194}
\bibinfo{author}{\bibfnamefont{R.~F.} \bibnamefont{Lebed}},
  \bibinfo{author}{\bibfnamefont{R.~E.} \bibnamefont{Mitchell}},
  \bibnamefont{and} \bibinfo{author}{\bibfnamefont{E.~S.}
  \bibnamefont{Swanson}}, \bibinfo{journal}{Prog. Part. Nucl. Phys.}
  \textbf{\bibinfo{volume}{93}}, \bibinfo{pages}{143} (\bibinfo{year}{2017}).

\bibitem[{\citenamefont{Guo et~al.}(2018)\citenamefont{Guo, Hanhart, Meißner,
  Wang, Zhao, and Zou}}]{2018-Guo-p15004-15004}
\bibinfo{author}{\bibfnamefont{F.-K.} \bibnamefont{Guo}},
  \bibinfo{author}{\bibfnamefont{C.}~\bibnamefont{Hanhart}},
  \bibinfo{author}{\bibfnamefont{U.-G.} \bibnamefont{Meißner}},
  \bibinfo{author}{\bibfnamefont{Q.}~\bibnamefont{Wang}},
  \bibinfo{author}{\bibfnamefont{Q.}~\bibnamefont{Zhao}}, \bibnamefont{and}
  \bibinfo{author}{\bibfnamefont{B.-S.} \bibnamefont{Zou}},
  \bibinfo{journal}{Rev. Mod. Phys.} \textbf{\bibinfo{volume}{90}},
  \bibinfo{pages}{015004} (\bibinfo{year}{2018}).

\bibitem[{\citenamefont{Olsen et~al.}(2018)\citenamefont{Olsen, Skwarnicki, and
  Zieminska}}]{2018-Olsen-p15003-15003}
\bibinfo{author}{\bibfnamefont{S.~L.} \bibnamefont{Olsen}},
  \bibinfo{author}{\bibfnamefont{T.}~\bibnamefont{Skwarnicki}},
  \bibnamefont{and}
  \bibinfo{author}{\bibfnamefont{D.}~\bibnamefont{Zieminska}},
  \bibinfo{journal}{Rev. Mod. Phys.} \textbf{\bibinfo{volume}{90}},
  \bibinfo{pages}{015003} (\bibinfo{year}{2018}).

\bibitem[{\citenamefont{Urey et~al.}(1932)\citenamefont{Urey, Brickwedde, and
  Murphy}}]{deuteron}
\bibinfo{author}{\bibfnamefont{H.~C.} \bibnamefont{Urey}},
  \bibinfo{author}{\bibfnamefont{F.~G.} \bibnamefont{Brickwedde}},
  \bibnamefont{and} \bibinfo{author}{\bibfnamefont{G.~M.}
  \bibnamefont{Murphy}}, \bibinfo{journal}{Phys. Rev.}
  \textbf{\bibinfo{volume}{40}}, \bibinfo{pages}{1} (\bibinfo{year}{1932}).
  
\bibitem{Bashkanov:2008ih} 
  M.~Bashkanov {\it et al.},
  Phys.\ Rev.\ Lett.\  {\bf 102}, 052301 (2009)
  
\bibitem[{\citenamefont{Adlarson et~al.}(2011)}]{2011-Adlarson-p242302-242302}
\bibinfo{author}{\bibfnamefont{P.}~\bibnamefont{Adlarson}} \bibnamefont{et~al.}
  (\bibinfo{collaboration}{WASA-at-COSY}), \bibinfo{journal}{Phys. Rev. Lett.}
  \textbf{\bibinfo{volume}{106}}, \bibinfo{pages}{242302}
  (\bibinfo{year}{2011}).

\bibitem[{\citenamefont{Adlarson et~al.}(2013)}]{2013-Adlarson-p229-236}
\bibinfo{author}{\bibfnamefont{P.}~\bibnamefont{Adlarson}} \bibnamefont{et~al.}
  (\bibinfo{collaboration}{WASA-at-COSY}), \bibinfo{journal}{Phys. Lett.}
  \textbf{\bibinfo{volume}{B721}}, \bibinfo{pages}{229} (\bibinfo{year}{2013}).

\bibitem[{\citenamefont{Adlarson et~al.}(2014)}]{2014-Adlarson-p202301-202301}
\bibinfo{author}{\bibfnamefont{P.}~\bibnamefont{Adlarson}} \bibnamefont{et~al.}
  (\bibinfo{collaboration}{WASA-at-COSY}), \bibinfo{journal}{Phys. Rev. Lett.}
  \textbf{\bibinfo{volume}{112}}, \bibinfo{pages}{202301}
  (\bibinfo{year}{2014}).

\bibitem[{\citenamefont{Adlarson et~al.}(2015)}]{2015-Adlarson-p325-332}
\bibinfo{author}{\bibfnamefont{P.}~\bibnamefont{Adlarson}} \bibnamefont{et~al.}
  (\bibinfo{collaboration}{WASA-at-COSY}), \bibinfo{journal}{Phys. Lett.}
  \textbf{\bibinfo{volume}{B743}}, \bibinfo{pages}{325} (\bibinfo{year}{2015}).
  
\bibitem[{\citenamefont{Yuan et~al.}(1999)\citenamefont{Yuan, Zhang, Yu, and
  Shen}}]{1999-Yuan-p45203-45203}
\bibinfo{author}{\bibfnamefont{X.~Q.} \bibnamefont{Yuan}},
  \bibinfo{author}{\bibfnamefont{Z.~Y.} \bibnamefont{Zhang}},
  \bibinfo{author}{\bibfnamefont{Y.~W.} \bibnamefont{Yu}}, \bibnamefont{and}
  \bibinfo{author}{\bibfnamefont{P.~N.} \bibnamefont{Shen}},
  \bibinfo{journal}{Phys. Rev.} \textbf{\bibinfo{volume}{C60}},
  \bibinfo{pages}{045203} (\bibinfo{year}{1999}).

\bibitem[{\citenamefont{Gal and Garcilazo}(2013)}]{2013-Gal-p172301-172301}
\bibinfo{author}{\bibfnamefont{A.}~\bibnamefont{Gal}} \bibnamefont{and}
  \bibinfo{author}{\bibfnamefont{H.}~\bibnamefont{Garcilazo}},
  \bibinfo{journal}{Phys. Rev. Lett.} \textbf{\bibinfo{volume}{111}},
  \bibinfo{pages}{172301} (\bibinfo{year}{2013}).

\bibitem[{\citenamefont{Chen et~al.}(2015{\natexlab{a}})\citenamefont{Chen,
  Cui, Chen, Steele, and Zhu}}]{2015-Chen-p25204-25204}
\bibinfo{author}{\bibfnamefont{H.-X.} \bibnamefont{Chen}},
  \bibinfo{author}{\bibfnamefont{E.-L.} \bibnamefont{Cui}},
  \bibinfo{author}{\bibfnamefont{W.}~\bibnamefont{Chen}},
  \bibinfo{author}{\bibfnamefont{T.}~\bibnamefont{Steele}}, \bibnamefont{and}
  \bibinfo{author}{\bibfnamefont{S.-L.} \bibnamefont{Zhu}},
  \bibinfo{journal}{Phys.Rev.} \textbf{\bibinfo{volume}{C91}},
  \bibinfo{pages}{025204} (\bibinfo{year}{2015}{\natexlab{a}}).

\bibitem[{\citenamefont{Huang et~al.}(2014)\citenamefont{Huang, Ping, and
  Wang}}]{2014-Huang-p34001-34001}
\bibinfo{author}{\bibfnamefont{H.}~\bibnamefont{Huang}},
  \bibinfo{author}{\bibfnamefont{J.}~\bibnamefont{Ping}}, \bibnamefont{and}
  \bibinfo{author}{\bibfnamefont{F.}~\bibnamefont{Wang}},
  \bibinfo{journal}{Phys. Rev.} \textbf{\bibinfo{volume}{C89}},
  \bibinfo{pages}{034001} (\bibinfo{year}{2014}).

\bibitem[{\citenamefont{Huang et~al.}(2015)\citenamefont{Huang, Zhang, Shen,
  and Wang}}]{2015-Huang-p71001-71001}
\bibinfo{author}{\bibfnamefont{F.}~\bibnamefont{Huang}},
  \bibinfo{author}{\bibfnamefont{Z.-Y.} \bibnamefont{Zhang}},
  \bibinfo{author}{\bibfnamefont{P.-N.} \bibnamefont{Shen}}, \bibnamefont{and}
  \bibinfo{author}{\bibfnamefont{W.-L.} \bibnamefont{Wang}},
  \bibinfo{journal}{Chin. Phys.} \textbf{\bibinfo{volume}{C39}},
  \bibinfo{pages}{071001} (\bibinfo{year}{2015}).



\bibitem[{\citenamefont{Dyson and Xuong}(1964)}]{dibaryon-sextet}
\bibinfo{author}{\bibfnamefont{F.~J.} \bibnamefont{Dyson}} \bibnamefont{and}
  \bibinfo{author}{\bibfnamefont{N.-H.} \bibnamefont{Xuong}},
  \bibinfo{journal}{Phys. Rev. Lett.} \textbf{\bibinfo{volume}{13}},
  \bibinfo{pages}{815} (\bibinfo{year}{1964}).
 
\bibitem{Molina:2021bwp}
R.~Molina, N.~Ikeno and E.~Oset,
[arXiv:2102.05575 [nucl-th]].

\bibitem{Ikeno:2021frl}
N.~Ikeno, R.~Molina and E.~Oset,
[arXiv:2103.01712 [nucl-th]].

\bibitem[{\citenamefont{Jaffe}(1977)}]{H-dibaryon}
\bibinfo{author}{\bibfnamefont{R.~L.} \bibnamefont{Jaffe}},
  \bibinfo{journal}{Phys. Rev. Lett.} \textbf{\bibinfo{volume}{38}},
  \bibinfo{pages}{195} (\bibinfo{year}{1977}).
  
%
%
%

\bibitem[{\citenamefont{Clement}(2017)}]{dibaryon-review}
\bibinfo{author}{\bibfnamefont{H.}~\bibnamefont{Clement}},
  \bibinfo{journal}{Prog. Part. Nucl. Phys.} 
  \textbf{\bibinfo{volume}{93}}, \bibinfo{pages}{195 } (\bibinfo{year}{2017}),
  ISSN \bibinfo{issn}{0146-6410}.
 

\bibitem{Goldman:1987ma}
J.~T.~Goldman, K.~Maltman, G.~J.~Stephenson, Jr., K.~E.~Schmidt and F.~Wang,
Phys. Rev. Lett. \textbf{59} (1987), 627
  
\bibitem{Li:1999bc} 
  Q.~B.~Li and P.~N.~Shen,
  Eur.\ Phys.\ J.\ A {\bf 8}, 417 (2000)
  
\bibitem{Pang:2003ty}
H.~R.~Pang, J.~L.~Ping, F.~Wang, J.~T.~Goldman and E.~G.~Zhao,
Phys. Rev. C \textbf{69} (2004), 065207

\bibitem{Chen:2011zzb}
M.~Chen, H.~Huang, J.~Ping and F.~Wang,
Phys. Rev. C \textbf{83} (2011), 015202

\bibitem{Sekihara:2018tsb}
T.~Sekihara, Y.~Kamiya and T.~Hyodo,
Phys. Rev. C \textbf{98} (2018) no.1, 015205

\bibitem{Xiao:2020alj}
C.~J.~Xiao, Y.~B.~Dong, T.~Gutsche, V.~E.~Lyubovitskij and D.~Y.~Chen,
Phys. Rev. D \textbf{101} (2020), 114032

\bibitem{Etminan:2014tya} 
  F.~Etminan {\it et al.} [HAL QCD Collaboration],
  Nucl.\ Phys.\ A {\bf 928}, 89 (2014)
  
\bibitem{Iritani:2018sra} 
  T.~Iritani {\it et al.} [HAL QCD Collaboration],
  Phys.\ Lett.\ B {\bf 792}, 284 (2019)
  
\bibitem{Morita:2016auo}
K.~Morita, A.~Ohnishi, F.~Etminan and T.~Hatsuda,
Phys. Rev. C \textbf{94} (2016) no.3, 031901
[erratum: Phys. Rev. C \textbf{100} (2019) no.6, 069902]

\bibitem{STAR:2018uho} 
  J.~Adam {\it et al.} [STAR Collaboration],
  Phys.\ Lett.\ B {\bf 790}, 490 (2019)
  
\bibitem{Chen:2019vdh}
X.~H.~Chen, Q.~N.~Wang, W.~Chen and H.~X.~Chen,
Chin. Phys. C \textbf{45} (2021) no.4, 041002

\bibitem{Shifman:1978bx} 
  M.~A.~Shifman, A.~I.~Vainshtein and V.~I.~Zakharov,
  Nucl.\ Phys.\ B {\bf 147}, 385 (1979).

\bibitem[{\citenamefont{Reinders et~al.}(1985)\citenamefont{Reinders,
  Rubinstein, and Yazaki}}]{QCD-sum-rule-Reinders}
\bibinfo{author}{\bibfnamefont{L.~J.} \bibnamefont{Reinders}},
  \bibinfo{author}{\bibfnamefont{H.}~\bibnamefont{Rubinstein}},
  \bibnamefont{and} \bibinfo{author}{\bibfnamefont{S.}~\bibnamefont{Yazaki}},
  \bibinfo{journal}{Phys. Rept.} \textbf{\bibinfo{volume}{127}},
  \bibinfo{pages}{1} (\bibinfo{year}{1985}).

\bibitem[{\citenamefont{Ioffe}(1981)}]{1981-Ioffe-p317-341}
\bibinfo{author}{\bibfnamefont{B.}~\bibnamefont{Ioffe}},
  \bibinfo{journal}{Nucl.Phys.} \textbf{\bibinfo{volume}{B188}},
  \bibinfo{pages}{317} (\bibinfo{year}{1981}).

\bibitem[{\citenamefont{Ioffe}(1983)}]{1983-Ioffe-p67-67}
\bibinfo{author}{\bibfnamefont{B.}~\bibnamefont{Ioffe}},
  \bibinfo{journal}{Z.Phys.} \textbf{\bibinfo{volume}{C18}},
  \bibinfo{pages}{67} (\bibinfo{year}{1983}).

\bibitem[{\citenamefont{Chen et~al.}(2014)\citenamefont{Chen, Cai, and
  Zhu}}]{2014-Chen-p201-215}
\bibinfo{author}{\bibfnamefont{W.}~\bibnamefont{Chen}},
  \bibinfo{author}{\bibfnamefont{Z.-X.} \bibnamefont{Cai}}, \bibnamefont{and}
  \bibinfo{author}{\bibfnamefont{S.-L.} \bibnamefont{Zhu}},
  \bibinfo{journal}{Nucl.Phys.} \textbf{\bibinfo{volume}{B887}},
  \bibinfo{pages}{201} (\bibinfo{year}{2014}).

\bibitem[{\citenamefont{Chen et~al.}(2017{\natexlab{a}})\citenamefont{Chen,
  Chen, Liu, Steele, and Zhu}}]{2017-Chen-p114005-114005}
\bibinfo{author}{\bibfnamefont{W.}~\bibnamefont{Chen}},
  \bibinfo{author}{\bibfnamefont{H.-X.} \bibnamefont{Chen}},
  \bibinfo{author}{\bibfnamefont{X.}~\bibnamefont{Liu}},
  \bibinfo{author}{\bibfnamefont{T.~G.} \bibnamefont{Steele}},
  \bibnamefont{and} \bibinfo{author}{\bibfnamefont{S.-L.} \bibnamefont{Zhu}},
  \bibinfo{journal}{Phys. Rev.} \textbf{\bibinfo{volume}{D95}},
  \bibinfo{pages}{114005} (\bibinfo{year}{2017}{\natexlab{a}}).

\bibitem[{\citenamefont{Chen et~al.}(2017{\natexlab{b}})\citenamefont{Chen,
  Chen, Liu, Steele, and Zhu}}]{2017-Chen-p114017-114017}
\bibinfo{author}{\bibfnamefont{W.}~\bibnamefont{Chen}},
  \bibinfo{author}{\bibfnamefont{H.-X.} \bibnamefont{Chen}},
  \bibinfo{author}{\bibfnamefont{X.}~\bibnamefont{Liu}},
  \bibinfo{author}{\bibfnamefont{T.~G.} \bibnamefont{Steele}},
  \bibnamefont{and} \bibinfo{author}{\bibfnamefont{S.-L.} \bibnamefont{Zhu}},
  \bibinfo{journal}{Phys. Rev.} \textbf{\bibinfo{volume}{D96}},
  \bibinfo{pages}{114017} (\bibinfo{year}{2017}{\natexlab{b}}).


\bibitem[{\citenamefont{Tanabashi
  et~al.}(2018{\natexlab{b}})\citenamefont{Tanabashi, Hagiwara, Hikasa,
  Nakamura, Sumino, Takahashi, Tanaka, Agashe, Aielli, Amsler et~al.}}]{PDG}
\bibinfo{author}{\bibfnamefont{M.}~\bibnamefont{Tanabashi}},
  \bibnamefont{et~al.} (\bibinfo{collaboration}{Particle Data Group}),
  \bibinfo{journal}{Phys. Rev. D} \textbf{\bibinfo{volume}{98}},
  \bibinfo{pages}{030001} (\bibinfo{year}{2018}{\natexlab{b}}).

\bibitem[{\citenamefont{Yang et~al.}(1993)\citenamefont{Yang, Hwang, Henley,
  and Kisslinger}}]{condensates1}
\bibinfo{author}{\bibfnamefont{K.-C.} \bibnamefont{Yang}},
  \bibinfo{author}{\bibfnamefont{W.-Y.~P.} \bibnamefont{Hwang}},
  \bibinfo{author}{\bibfnamefont{E.~M.} \bibnamefont{Henley}},
  \bibnamefont{and} \bibinfo{author}{\bibfnamefont{L.~S.}
  \bibnamefont{Kisslinger}}, \bibinfo{journal}{Phys. Rev. D}
  \textbf{\bibinfo{volume}{47}}, \bibinfo{pages}{3001} (\bibinfo{year}{1993}).

\bibitem[{\citenamefont{Jamin}(2002)}]{condensates2}
\bibinfo{author}{\bibfnamefont{M.}~\bibnamefont{Jamin}},
  \bibinfo{journal}{Physics Letters B} \textbf{\bibinfo{volume}{538}},
  \bibinfo{pages}{71 } (\bibinfo{year}{2002}), ISSN \bibinfo{issn}{0370-2693}.

\bibitem[{\citenamefont{Gimenez et~al.}(2005)\citenamefont{Gimenez, Lubicz,
  Mescia, Porretti, and Reyes}}]{condensates3}
\bibinfo{author}{\bibfnamefont{V.}~\bibnamefont{Gimenez}},
  \bibinfo{author}{\bibfnamefont{V.}~\bibnamefont{Lubicz}},
  \bibinfo{author}{\bibfnamefont{F.}~\bibnamefont{Mescia}},
  \bibinfo{author}{\bibfnamefont{V.}~\bibnamefont{Porretti}}, \bibnamefont{and}
  \bibinfo{author}{\bibfnamefont{J.}~\bibnamefont{Reyes}},
  \bibinfo{journal}{Eur. Phys. J.} \textbf{\bibinfo{volume}{C41}},
  \bibinfo{pages}{535} (\bibinfo{year}{2005}).

\bibitem[{\citenamefont{Ioffe and Zyablyuk}(2003)}]{condensates4}
\bibinfo{author}{\bibfnamefont{B.~L.} \bibnamefont{Ioffe}} \bibnamefont{and}
  \bibinfo{author}{\bibfnamefont{K.~N.} \bibnamefont{Zyablyuk}},
  \bibinfo{journal}{Eur. Phys. J.} \textbf{\bibinfo{volume}{C27}},
  \bibinfo{pages}{229} (\bibinfo{year}{2003}).

\bibitem[{\citenamefont{Ovchinnikov and Pivovarov}(1988)}]{condensates5}
\bibinfo{author}{\bibfnamefont{A.~A.} \bibnamefont{Ovchinnikov}}
  \bibnamefont{and} \bibinfo{author}{\bibfnamefont{A.~A.}
  \bibnamefont{Pivovarov}}, \bibinfo{journal}{Sov. J. Nucl. Phys.}
  \textbf{\bibinfo{volume}{48}}, \bibinfo{pages}{721} (\bibinfo{year}{1988}),
  \bibinfo{note}{[Yad. Fiz.48,1135(1988)]}.

\bibitem[{\citenamefont{Chen et~al.}(2008)\citenamefont{Chen, Hosaka, and
  Zhu}}]{condensates-chenhuaxing}
\bibinfo{author}{\bibfnamefont{H.-X.} \bibnamefont{Chen}},
  \bibinfo{author}{\bibfnamefont{A.}~\bibnamefont{Hosaka}}, \bibnamefont{and}
  \bibinfo{author}{\bibfnamefont{S.-L.} \bibnamefont{Zhu}},
  \bibinfo{journal}{Phys. Rev. D} \textbf{\bibinfo{volume}{78}},
  \bibinfo{pages}{054017} (\bibinfo{year}{2008}).

\bibitem[{\citenamefont{Chen et~al.}(2015{\natexlab{b}})\citenamefont{Chen,
  Cui, Chen, Steele, and Zhu}}]{QCD-sum-rule-chenhuaxing}
\bibinfo{author}{\bibfnamefont{H.-X.} \bibnamefont{Chen}},
  \bibinfo{author}{\bibfnamefont{E.-L.} \bibnamefont{Cui}},
  \bibinfo{author}{\bibfnamefont{W.}~\bibnamefont{Chen}},
  \bibinfo{author}{\bibfnamefont{T.~G.} \bibnamefont{Steele}},
  \bibnamefont{and} \bibinfo{author}{\bibfnamefont{S.-L.} \bibnamefont{Zhu}},
  \bibinfo{journal}{Phys. Rev. C} \textbf{\bibinfo{volume}{91}},
  \bibinfo{pages}{025204} (\bibinfo{year}{2015}{\natexlab{b}}).
  
\bibitem{Cho:2010db}
S.~Cho \textit{et al.} [ExHIC],
Phys. Rev. Lett. \textbf{106} (2011), 212001

\bibitem{Cho:2011ew}
S.~Cho \textit{et al.} [ExHIC],
Phys. Rev. C \textbf{84} (2011), 064910

\bibitem{Yuan:2021yks}
C.~Z.~Yuan and M.~Karliner,
[arXiv:2103.06658 [hep-ex]].

\end{thebibliography}

\begin{widetext}
\appendix
\section{The spectral densities  \label{spectraldensities}}
We calculate the spectral densities for both the $J^P=1^+$ axial-vector channel and $J^P=2^+$ tensor channel up to dimension-16 condensates by using two interpolating currents $J^{N\Omega}_{\mu\nu,\,1}(x)$ and $J^{N\Omega}_{\mu\nu,\,2}(x)$, and collect all of them below:	
\begin{itemize}
\item
For the $J^P=2^+$ tensor channel in $J^{N\Omega}_{\mu\nu,\,1}(x)$:

\begin{align}
\begin{autobreak}
	\rho(s) =
	\frac{s^7}{3^2 5^2 7^2 2^{15} \pi ^{10}}
 -\frac{43 \dss \ms s^5}{105\times 2^{14} \pi ^8}
 -\frac{\dqq \ms s^5}{105\times 2^{10} \pi ^8}
 -\frac{\dGG s^5}{35\times 9\times 2^{17} \pi ^{10}}
 +\frac{5 \dss^2 s^4}{3^3 2^8 \pi ^6}
 +\frac{5 \dqq \dss s^4}{3^3 2^9 \pi ^6}
 +\frac{5 \dqq^2 s^4}{63\times 2^9 \pi ^6}
 -\frac{25 \dsGs \ms s^4}{27\times 7\times 2^9 \pi ^8}
 +\frac{7 \dsGs \dss s^3}{3^3 2^5 \pi ^6}
 +\frac{\dqq \dsGs s^3}{3^3 2^4 \pi ^6}
 +\frac{5 \dqGq \dqq s^3}{3^3 2^8 \pi ^6}
 -\frac{83 \dGG \dss \ms s^3}{3^4 2^{11} \pi ^8}
 +\frac{7 \dGG \dqq \ms s^3}{3^3 2^{11} \pi ^8}
 +\frac{5 \dss^3 \ms s^2}{48 \pi ^4}
 -\frac{5 \dqq \dss^2 \ms s^2}{48 \pi ^4}
 -\frac{7 \dqq^2 \dss \ms s^2}{48 \pi ^4}
 -\frac{5 \dqq^3 \ms s^2}{24 \pi ^4}
 +\frac{25 \dGG \dss^2 s^2}{3^2 2^9 \pi ^6} 
 -\frac{5 \dGG \dqq \dss s^2}{3^3 2^7 \pi ^6} 
 +\frac{5 \dsGs^2 s^2}{3\times 2^7 \pi ^6}
 +\frac{5 \dqGq \dsGs s^2}{3\times 2^{10} \pi ^6}
 -\frac{\dGG \dqq^2 s^2}{3^2 2^8 \pi ^6} 
 -\frac{5 \dGG \dsGs \ms s^2}{3^2 2^9 \pi ^8}
 +\frac{35 \dGG \dqGq \ms s^2}{3^2 2^{13} \pi ^8}
 +\frac{5 \dqq \dss^3 s}{9 \pi ^2}
 +\frac{20 \dqq^2 \dss^2 s}{27 \pi ^2} 
 +\frac{40 \dqq^3 \dss s}{27 \pi ^2}
 +\frac{10 \dsGs \dss^2 \ms s}{27 \pi ^4}
 -\frac{10 \dqq \dsGs \dss \ms s}{27 \pi ^4}
 -\frac{115 \dqGq \dqq \dss \ms s}{3^2 2^6 \pi ^4}
 -\frac{235 \dqq^2 \dsGs \ms s}{3^3 2^5 \pi ^4}
 -\frac{5 \dqGq \dqq^2 \ms s}{6 \pi ^4}
 +\frac{125 \dGG \dsGs \dss s}{3^4 2^7 \pi ^6}
 -\frac{25 \dGG \dqGq \dss s}{3^3 2^8 \pi ^6}
 -\frac{55 \dGG \dqq \dsGs s}{3^4 2^8 \pi ^6}
 -\frac{115 \dGG \dqGq \dqq s}{3^3 2^{11} \pi ^6}
 -\frac{5 \dGG^2 \dss \ms s}{3^3 2^{10} \pi ^8}
 +\frac{25 \dGG^2 \dqq \ms s}{3^4 2^{12} \pi ^8}
 +\frac{5 \dqq \dsGs \dss^2}{6 \pi ^2}
 +\frac{5 \dqGq \dqq \dss^2}{18 \pi ^2}
 +\frac{35 \dqq^2 \dsGs \dss}{54 \pi ^2}
 +\frac{5 \dqGq \dqq^2 \dss}{3 \pi ^2} 
 +\frac{20 \dqq^3 \dsGs}{27 \pi ^2}
 +\frac{25 \dGG \dss^3 \ms}{3^3 2^5 \pi ^4}
 -\frac{5 \dGG \dqq \dss^2 \ms}{3^3 2^5 \pi ^4}
 +\frac{35 \dsGs^2 \dss \ms}{3^2 2^5 \pi ^4}
 +\frac{5 \dqGq \dsGs \dss \ms}{3^2 2^6 \pi ^4}
 -\frac{175 \dGG \dqq^2 \dss \ms}{3^4 2^6 \pi ^4}
 -\frac{25 \dqq \dsGs^2 \ms}{6^3 \pi ^4}
 -\frac{35 \dqGq \dqq \dsGs \ms}{3^2 2^5 \pi ^4}
 -\frac{5 \dqGq^2 \dqq \ms}{16 \pi ^4}
 +\frac{35 \dGG^2 \dss^2}{3^4 2^{10} \pi ^6}
 -\frac{55 \dGG^2 \dqq \dss}{3^4 2^{11} \pi ^6}
 +\frac{5 \dGG \dsGs^2}{3\times 2^{10} \pi ^6} 
 -\frac{5 \dGG \dqGq \dsGs}{3\times 2^{10} \pi ^6}
 -\frac{5 \dGG^2 \dqq^2}{3^3 2^{11} \pi ^6}
 -\frac{25 \dGG \dqGq^2}{3^3 2^{11} \pi ^6}
 +\frac{5 \dGG \dqq \dss^3 \delta (s)}{2^3 3^4 \pi ^2}
 +\frac{5 \dqq \dsGs^2 \dss \delta (s)}{24 \pi ^2}
 -\frac{5 \dGG \dqq^3 \dss \delta (s)}{81 \pi ^2}
 +\frac{5 \dGG \dqGq \dss^2 \ms \delta (s)}{3\times 2^7 \pi ^4}
 +\frac{5 \dGG \dqq \dsGs \dss \ms \delta (s)}{3^2 2^6 \pi ^4}
 +\frac{5 \dGG \dqGq \dqq^2 \ms \delta (s)}{3^2 2^4 \pi ^4}
 -\frac{5 \dGG^2 \dqGq \dss \delta (s)}{3^4 2^{11} \pi ^6}
 -\frac{5 \dGG^2 \dqq \dsGs \delta (s)}{3^4 2^{11} \pi ^6}
 -\frac{80 \dqq^3 \dss^2 \ms \delta (s)}{9}
\end{autobreak}
\end{align}

\item
	
	For the $J^P=1^+$ axial-vector channel in $J^{N\Omega}_{\mu\nu,\,1}(x)$:
	
\begin{align}
\begin{autobreak}
	\rho(s)=
	\frac{s^7}{7\times 5^2 3^3 2^{18} \pi ^{10}}
	-\frac{\dss \ms s^5}{105\times 2^{13} \pi ^8}
	-\frac{\dqq \ms s^5}{75\times 2^{13} \pi ^8}
	+\frac{\dGG s^5}{35\times 9\times 2^{17} \pi ^{10}}
	+\frac{\dss^2 s^4}{45\times 2^8 \pi ^6}
	+\frac{\dqq \dss s^4}{15\times 2^{10} \pi ^6}
	+\frac{\dqq^2 s^4}{45\times 2^9 \pi ^6}
	-\frac{\dsGs \ms s^4}{45\times 2^{10} \pi ^8}
	+\frac{7 \dsGs \dss s^3}{15\times 2^9 \pi ^6}
	+\frac{\dqq \dsGs s^3}{3^2 2^8 \pi ^6}
	+\frac{\dqGq \dqq s^3}{5\times 2^{10} \pi ^6}
	-\frac{\dGG \dss \ms s^3}{15\times 2^{11} \pi ^8}
	+\frac{5 \dGG \dqq \ms s^3}{3^2 2^{14} \pi ^8}
	+\frac{\dss^3 \ms s^2}{96 \pi ^4}
	-\frac{\dqq \dss^2 \ms s^2}{24 \pi ^4}
	-\frac{\dqq^2 \dss \ms s^2}{192 \pi ^4}
	-\frac{\dqq^3 \ms s^2}{24 \pi ^4}
	+\frac{5 \dGG \dss^2 s^2}{3^2 2^{10} \pi ^6}
	-\frac{\dGG \dqq \dss s^2}{3^2 2^8 \pi ^6}
	+\frac{\dsGs^2 s^2}{3\times 2^8 \pi ^6}
	+\frac{\dqGq \dsGs s^2}{3\times 2^{10} \pi ^6}
	+\frac{\dGG \dqq^2 s^2}{3^2 2^{11} \pi ^6}
	-\frac{5 \dGG \dsGs \ms s^2}{3^2 2^{13} \pi ^8}
	+\frac{\dGG \dqGq \ms s^2}{2^{13} \pi ^8}
	+\frac{\dqq \dss^3 s}{6 \pi ^2}
	+\frac{\dqq^2 \dss^2 s}{18 \pi ^2}
	+\frac{\dqq^3 \dss s}{3 \pi ^2}
	+\frac{\dsGs \dss^2 \ms s}{36 \pi ^4}
	-\frac{25 \dqq \dsGs \dss \ms s}{192 \pi ^4}
	+\frac{\dqGq \dqq \dss \ms s}{192 \pi ^4}
	-\frac{\dqq^2 \dsGs \ms s}{72 \pi ^4}
	-\frac{3 \dqGq \dqq^2 \ms s}{16 \pi ^4}
	+\frac{25 \dGG \dsGs \dss s}{3^3 2^{10} \pi ^6}
	+\frac{\dGG \dqGq \dss s}{2^{10} \pi ^6}
	-\frac{7 \dGG \dqq \dsGs s}{3^2 2^{10} \pi ^6}
	-\frac{\dGG \dqGq \dqq s}{3^2 2^{11} \pi ^6}
	-\frac{\dGG^2 \dss \ms s}{3^2 2^{13} \pi ^8}
	+\frac{\dGG^2 \dqq \ms s}{3\times 2^{14} \pi ^8}
	+\frac{\dqq \dsGs \dss^2}{4 \pi ^2}
	+\frac{\dqGq \dqq^2 \dss}{2 \pi ^2}
	+\frac{2 \dqq^3 \dsGs}{9 \pi ^2}
	-\frac{\dGG \dqq \dss^2 \ms}{3^2 2^6 \pi ^4}
	+\frac{\dqGq \dsGs \dss \ms}{3\times 2^7 \pi ^4}
	+\frac{5 \dqq \dsGs^2 \ms}{144 \pi ^4}
	-\frac{3 \dqGq^2 \dqq \ms}{32 \pi ^4}
	-\frac{11 \dGG^2 \dqq \dss}{3^3 2^{12} \pi ^6}
	-\frac{\dGG \dqGq \dsGs}{2^{11} \pi ^6}
	+\frac{\dGG \dqq \dss^3 \delta (s)}{3^3 2^4 \pi ^2}
	-\frac{\dGG \dqq^2 \dss^2 \delta (s)}{108 \pi ^2}
	+\frac{\dqq \dsGs^2 \dss \delta (s)}{16 \pi ^2}
	-\frac{\dqGq \dqq \dsGs \dss \delta (s)}{18 \pi ^2}
	+\frac{\dGG \dqq^3 \dss \delta (s)}{6^3 \pi ^2}
	+\frac{\dqGq^2 \dqq \dss \delta (s)}{8 \pi ^2}
	-\frac{\dqq^2 \dsGs^2 \delta (s)}{36 \pi ^2}
	+\frac{\dqGq \dqq^2 \dsGs \delta (s)}{6 \pi ^2}
	-\frac{17 \dGG \dsGs \dss^2 \ms \delta (s)}{3^3 2^7 \pi ^4}
	+\frac{\dGG \dqGq \dss^2 \ms \delta (s)}{384 \pi ^4}
	+\frac{7 \dGG \dqq \dsGs \dss \ms \delta (s)}{3^2 2^9 \pi ^4}
	+\frac{7 \dGG \dqGq \dqq \dss \ms \delta (s)}{3^2 2^9 \pi ^4}
	-\frac{\dsGs^3 \ms \delta (s)}{3\times 2^7 \pi ^4}
	+\frac{\dqGq \dsGs^2 \ms \delta (s)}{3^2 2^8 \pi ^4}
	+\frac{\dGG \dqq^2 \dsGs \ms \delta (s)}{3\times 2^8 \pi ^4}
	+\frac{\dqGq^2 \dsGs \ms \delta (s)}{3\times 2^8 \pi ^4}
	+\frac{\dGG \dqGq \dqq^2 \ms \delta (s)}{3\times 2^7 \pi ^4}
	-\frac{\dqGq^3 \ms \delta (s)}{128 \pi ^4}
	-\frac{\dGG^2 \dsGs \dss \delta (s)}{3^2 2^{12} \pi ^6}
	-\frac{\dGG^2 \dqGq \dss \delta (s)}{3^2 2^{13} \pi ^6}
	-\frac{\dGG^2 \dqq \dsGs \delta (s)}{3^2 2^{13} \pi ^6}
	+\frac{\dGG^2 \dqGq \dqq \delta (s)}{3^3 2^{13} \pi ^6}
	-\frac{2 \dqq^2 \dss^3 \ms \delta (s)}{9}
	-\frac{16 \dqq^3 \dss^2 \ms \delta (s)}{9}
\end{autobreak}
\end{align}

\item
	
	For the $J^P=2^+$ tensor channel in $J^{N\Omega}_{\mu\nu,\,2}(x)$:
	
\begin{align}
\begin{autobreak}
	\rho(s)=
	\frac{s^7}{45\times7^2 2^{20} \pi ^{10}}
	 -\frac{43 \dss \ms s^5}{21\times2^{19} \pi ^8}
	 -\frac{\dqq \ms s^5}{15\times2^{14} \pi ^8}
	 -\frac{\dGG s^5}{63\times2^{22} \pi ^{10}}
	 +\frac{25 \dss^2 s^4}{3^3 2^{13} \pi ^6}
	 +\frac{35 \dqq \dss s^4}{3^3 2^{13} \pi ^6}
	 +\frac{5 \dqq^2 s^4}{3^2 2^{13} \pi ^6}
	 -\frac{125 \dsGs \ms s^4}{7\times3^3 2^{14} \pi ^8}
	 -\frac{5 \dqGq \ms s^4}{3\times2^{16} \pi ^8}
	 +\frac{35 \dsGs \dss s^3}{3^3 2^{10} \pi ^6}
	 +\frac{\dqGq \dss s^3}{3\times2^9 \pi ^6}
	 +\frac{7 \dqq \dsGs s^3}{3^3 2^8 \pi ^6}
	 +\frac{65 \dqGq \dqq s^3}{3^3 2^{12} \pi ^6}
	 -\frac{83\times5 \dGG \dss \ms s^3}{3^4 2^{16} \pi ^8}
	 -\frac{23 \dGG \dqq \ms s^3}{3^3 2^{15} \pi ^8}
	 +\frac{25 \dss^3 \ms s^2}{3\times2^9 \pi ^4}
	 -\frac{35 \dqq \dss^2 \ms s^2}{3\times2^8 \pi ^4}
	 -\frac{49 \dqq^2 \dss \ms s^2}{3\times2^8 \pi ^4}
	 -\frac{25 \dqq^3 \ms s^2}{3\times2^8 \pi ^4}
	 +\frac{125 \dGG \dss^2 s^2}{3^2 2^{14} \pi ^6}
	 +\frac{65 \dGG \dqq \dss s^2}{3^3 2^{12} \pi ^6}
	 +\frac{25 \dsGs^2 s^2}{3\times2^{12} \pi ^6}
	 +\frac{125 \dqGq \dsGs s^2}{3\times2^{14} \pi ^6}
	 -\frac{7 \dGG \dqq^2 s^2}{3^2 2^{12} \pi ^6}
	 +\frac{3 \dqGq^2 s^2}{2^{12} \pi ^6}
	 -\frac{25 \dGG \dsGs \ms s^2}{3^2 2^{14} \pi ^8}
	 -\frac{25 \dGG \dqGq \ms s^2}{3^2 2^{17} \pi ^8}
	 +\frac{35 \dqq \dss^3 s}{144 \pi ^2}
	 +\frac{35 \dqq^2 \dss^2 s}{108 \pi ^2}
	 +\frac{25 \dqq^3 \dss s}{108 \pi ^2}
	 +\frac{25 \dsGs \dss^2 \ms s}{3^3 2^4 \pi ^4}
	 -\frac{25 \dqGq \dss^2 \ms s}{3\times2^7 \pi ^4}
	 -\frac{35 \dqq \dsGs \dss \ms s}{6^3 \pi ^4}
	 -\frac{65\times23 \dqGq \dqq \dss \ms s}{3^2 2^{10} \pi ^4}
	 -\frac{47\times35 \dqq^2 \dsGs \ms s}{3^3 2^9 \pi ^4}
	 -\frac{25 \dqGq \dqq^2 \ms s}{192 \pi ^4}
	 +\frac{25^2 \dGG \dsGs \dss s}{3^4 2^{12} \pi ^6}
	 +\frac{35 \dGG \dqGq \dss s}{3^3 2^{12} \pi ^6}
	 +\frac{155 \dGG \dqq \dsGs s}{3^4 2^{12} \pi ^6}
	 -\frac{55\times13 \dGG \dqGq \dqq s}{3^3 2^{15} \pi ^6}
	 -\frac{25 \dGG^2 \dss \ms s}{3^3 2^{15} \pi ^8}
	 -\frac{5 \dGG^2 \dqq \ms s}{3^4 2^{16} \pi ^8}
	 +\frac{5 \dqGq \dss^3}{48 \pi ^2}
	 +\frac{35 \dqq \dsGs \dss^2}{96 \pi ^2}
	 +\frac{65 \dqGq \dqq \dss^2}{3^2 2^5 \pi ^2}
	 +\frac{49\times5 \dqq^2 \dsGs \dss}{3^3 2^5 \pi ^2}
	 +\frac{25 \dqGq \dqq^2 \dss}{96 \pi ^2}
	 +\frac{25 \dqq^3 \dsGs}{6^3 \pi ^2}
	 +\frac{125 \dGG \dss^3 \ms}{3^3 2^{10} \pi ^4}
	 -\frac{79\times5 \dGG \dqq \dss^2 \ms}{3^3 2^9 \pi ^4}
	 +\frac{175 \dsGs^2 \dss \ms}{3^2 2^{10} \pi ^4}
	 -\frac{149\times5 \dqGq \dsGs \dss \ms}{3^2 2^{10} \pi ^4}
	 -\frac{49\times25 \dGG \dqq^2 \dss \ms}{3^4 2^{10} \pi ^4}
	 -\frac{15 \dqGq^2 \dss \ms}{2^9 \pi ^4}
	 -\frac{175 \dqq \dsGs^2 \ms}{3^3 2^7 \pi ^4}
	 -\frac{79\times5 \dqGq \dqq \dsGs \ms}{3^2 2^9 \pi ^4}
	 -\frac{25 \dqGq^2 \dqq \ms}{2^9 \pi ^4}
	 +\frac{175 \dGG^2 \dss^2}{3^4 2^{15} \pi ^6}
	 -\frac{25 \dGG^2 \dqq \dss}{3^4 2^{15} \pi ^6}
	 +\frac{25 \dGG \dsGs^2}{3\times2^{15} \pi ^6}
	 -\frac{5 \dGG \dqGq \dsGs}{3\times2^{14} \pi ^6}
	 -\frac{35 \dGG^2 \dqq^2}{3^3 2^{15} \pi ^6}
	 -\frac{175 \dGG \dqGq^2}{3^3 2^{15} \pi ^6}
	 +\frac{43\times5 \dGG \dqq \dss^3 \delta (s)}{3^4 2^7 \pi ^2}
	 +\frac{5 \dqGq \dsGs \dss^2 \delta (s)}{64 \pi ^2}
	 +\frac{35 \dqq \dsGs^2 \dss \delta (s)}{3\times2^7 \pi ^2}
	 -\frac{25 \dGG \dqq^3 \dss \delta (s)}{3^4 2^5 \pi ^2}
	 -\frac{15 \dGG \dqGq \dss^2 \ms \delta (s)}{2^{11} \pi ^4}
	 -\frac{145 \dGG \dqq \dsGs \dss \ms \delta (s)}{3^2 2^{10} \pi ^4}
	 -\frac{15 \dqGq \dsGs^2 \ms \delta (s)}{2^{10} \pi ^4}
	 +\frac{25 \dGG \dqGq \dqq^2 \ms \delta (s)}{3^2 2^9 \pi ^4}
	 -\frac{35 \dGG^2 \dqGq \dss \delta (s)}{3^4 2^{15} \pi ^6}
	 -\frac{35 \dGG^2 \dqq \dsGs \delta (s)}{3^4 2^{15} \pi ^6}
	 -\frac{25 \dqq^3 \dss^2 \ms \delta (s)}{18}
\end{autobreak}
\end{align}

\item
	
	For the $J^P=1^+$ axial-vector channel in $J^{N\Omega}_{\mu\nu,\,2}(x)$:
	
\begin{align}
\begin{autobreak}
	\rho(s)=
	\frac{s^7}{35\ 3^3 2^{23} \pi ^{10}}
	 -\frac{\dss \ms s^5}{21\ 2^{18} \pi ^8}
	 -\frac{7 \dqq \ms s^5}{75\ 2^{17} \pi ^8}
	 +\frac{\dGG s^5}{63\ 2^{22} \pi ^{10}}
	 +\frac{\dss^2 s^4}{3^2 2^{13} \pi ^6}
	 +\frac{7 \dqq \dss s^4}{15\ 2^{14} \pi ^6}
	 +\frac{7 \dqq^2 s^4}{45\ 2^{13} \pi ^6}
	 -\frac{\dsGs \ms s^4}{3^2 2^{15} \pi ^8}
	 -\frac{3 \dqGq \ms s^4}{5\ 2^{17} \pi ^8}
	 +\frac{7 \dsGs \dss s^3}{3\ 2^{14} \pi ^6}
	 +\frac{\dqGq \dss s^3}{2^{13} \pi ^6}
	 +\frac{7 \dqq \dsGs s^3}{3^2 2^{12} \pi ^6}
	 +\frac{13 \dqGq \dqq s^3}{5\ 2^{14} \pi ^6}
	 -\frac{\dGG \dss \ms s^3}{3\ 2^{16} \pi ^8}
	 -\frac{\dGG \dqq \ms s^3}{3^2 2^{18} \pi ^8}
	 +\frac{5 \dss^3 \ms s^2}{3\ 2^{10} \pi ^4}
	 -\frac{7 \dqq \dss^2 \ms s^2}{3\ 2^7 \pi ^4}
	 -\frac{7 \dqq^2 \dss \ms s^2}{3\ 2^{10} \pi ^4}
	 -\frac{5 \dqq^3 \ms s^2}{3\ 2^8 \pi ^4}
	 +\frac{25 \dGG \dss^2 s^2}{3^2 2^{15} \pi ^6}
	 +\frac{\dGG \dqq \dss s^2}{3^2 2^{11} \pi ^6}
	 +\frac{5 \dsGs^2 s^2}{3\ 2^{13} \pi ^6}
	 +\frac{25 \dqGq \dsGs s^2}{3\ 2^{14} \pi ^6}
	 +\frac{7 \dGG \dqq^2 s^2}{3^2 2^{15} \pi ^6}
	 +\frac{3 \dqGq^2 s^2}{2^{14} \pi ^6}
	 -\frac{25 \dGG \dsGs \ms s^2}{3^2 2^{18} \pi ^8}
	 +\frac{\dGG \dqGq \ms s^2}{2^{17} \pi ^8}
	 +\frac{7 \dqq \dss^3 s}{96 \pi ^2}
	 +\frac{7 \dqq^2 \dss^2 s}{3^2 2^5 \pi ^2}
	 +\frac{5 \dqq^3 \dss s}{96 \pi ^2}
	 +\frac{5 \dsGs \dss^2 \ms s}{3^2 2^7 \pi ^4}
	 -\frac{3 \dqGq \dss^2 \ms s}{128 \pi ^4}
	 -\frac{175 \dqq \dsGs \dss \ms s}{3\ 2^{10} \pi ^4}
	 -\frac{13 \dqGq \dqq \dss \ms s}{3\ 2^{10} \pi ^4}
	 -\frac{7 \dqq^2 \dsGs \ms s}{3^2 2^7 \pi ^4}
	 -\frac{15 \dqGq \dqq^2 \ms s}{2^9 \pi ^4}
	 +\frac{125 \dGG \dsGs \dss s}{3^3 2^{15} \pi ^6}
	 +\frac{5 \dGG \dqq \dsGs s}{3^2 2^{14} \pi ^6}
	 -\frac{\dGG \dqGq \dqq s}{3^2 2^{15} \pi ^6}
	 -\frac{5 \dGG^2 \dss \ms s}{3^2 2^{18} \pi ^8}
	 +\frac{\dGG^2 \dqq \ms s}{3\ 2^{18} \pi ^8}
	 +\frac{\dqGq \dss^3}{32 \pi ^2}
	 +\frac{7 \dqq \dsGs \dss^2}{64 \pi ^2}
	 +\frac{5 \dqGq \dqq^2 \dss}{64 \pi ^2}
	 +\frac{5 \dqq^3 \dsGs}{144 \pi ^2}
	 -\frac{79 \dGG \dqq \dss^2 \ms}{3^2 2^{10} \pi ^4}
	 -\frac{149 \dqGq \dsGs \dss \ms}{3\ 2^{11} \pi ^4}
	 -\frac{35 \dqq \dsGs^2 \ms}{3^2 2^8 \pi ^4}
	 -\frac{15 \dqGq^2 \dqq \ms}{2^{10} \pi ^4}
	 -\frac{5 \dGG^2 \dqq \dss}{3^3 2^{16} \pi ^6}
	 -\frac{\dGG \dqGq \dsGs}{2^{15} \pi ^6}
	 +\frac{43 \dGG \dqq \dss^3 \delta (s)}{3^3 2^8 \pi ^2}
	 +\frac{3 \dqGq \dsGs \dss^2 \delta (s)}{128 \pi ^2}
	 -\frac{7 \dGG \dqq^2 \dss^2 \delta (s)}{3^3 2^6 \pi ^2}
	 -\frac{\dqGq^2 \dss^2 \delta (s)}{128 \pi ^2}
	 +\frac{7 \dqq \dsGs^2 \dss \delta (s)}{256 \pi ^2}
	 -\frac{23 \dqGq \dqq \dsGs \dss \delta (s)}{3^2 2^6 \pi ^2}
	 +\frac{5 \dGG \dqq^3 \dss \delta (s)}{3^3 2^8 \pi ^2}
	 +\frac{5 \dqGq^2 \dqq \dss \delta (s)}{256 \pi ^2}
	 -\frac{7 \dqq^2 \dsGs^2 \delta (s)}{3^2 2^6 \pi ^2}
	 +\frac{5 \dqGq \dqq^2 \dsGs \delta (s)}{192 \pi ^2}
	 -\frac{85 \dGG \dsGs \dss^2 \ms \delta (s)}{3^3 2^{12} \pi ^4}
	 -\frac{5 \dGG \dqGq \dss^2 \ms \delta (s)}{3\ 2^{10} \pi ^4}
	 -\frac{263 \dGG \dqq \dsGs \dss \ms \delta (s)}{3^2 2^{13} \pi ^4}
	 +\frac{79 \dGG \dqGq \dqq \dss \ms \delta (s)}{3^2 2^{13} \pi ^4}
	 -\frac{5 \dsGs^3 \ms \delta (s)}{3\ 2^{12} \pi ^4}
	 -\frac{119 \dqGq \dsGs^2 \ms \delta (s)}{3^2 2^{12} \pi ^4}
	 +\frac{7 \dGG \dqq^2 \dsGs \ms \delta (s)}{3\ 2^{12} \pi ^4}
	 +\frac{25 \dqGq^2 \dsGs \ms \delta (s)}{3\ 2^{12} \pi ^4}
	 +\frac{5 \dGG \dqGq \dqq^2 \ms \delta (s)}{3\ 2^{12} \pi ^4}
	 -\frac{5 \dqGq^3 \ms \delta (s)}{2^{12} \pi ^4}
	 -\frac{5 \dGG^2 \dsGs \dss \delta (s)}{3^2 2^{17} \pi ^6}
	 -\frac{\dGG^2 \dqGq \dss \delta (s)}{3^2 2^{17} \pi ^6}
	 -\frac{\dGG^2 \dqq \dsGs \delta (s)}{3^2 2^{17} \pi ^6}
	 +\frac{7 \dGG^2 \dqGq \dqq \delta (s)}{3^3 2^{17} \pi ^6}
	 -\frac{7 \dqq^2 \dss^3 \ms \delta (s)}{72}
	 -\frac{5 \dqq^3 \dss^2 \ms \delta (s)}{18}
\end{autobreak}
\end{align}

\end{itemize}

\end{widetext}

\end{document}